\documentclass[10pt,twocolumn]{article}

\usepackage[letterpaper,margin=0.75in,columnsep=0.25in]{geometry}
\usepackage{newtxtext,newtxmath}
\usepackage{graphicx}
\usepackage{booktabs}
\usepackage{makecell}
\usepackage{caption}
\usepackage{microtype}
\usepackage{enumitem}
\usepackage{url}
\usepackage[hidelinks,breaklinks]{hyperref}
\setlist{itemsep=2pt,topsep=3pt,parsep=0pt}

\title{\bfseries Re-ranking and Late Interaction Drive Retrieval Quality: A Controlled Comparison of RAG Strategies for Scientific Question Answering}
\author{Bhagyesh Rathi$^{1}$, Eshan Chawla$^{1}$, William B.~Andreopoulos$^{1}$\\[4pt]
\normalsize $^{1}$San Jose State University, Department of Computer Science}
\date{}

\begin{document}

\twocolumn[
  \maketitle
  \vspace{-2em}
]

\noindent\textbf{\textit{Abstract}--- Retrieval-Augmented Generation (RAG) is
now the standard way to ground Large Language Models (LLMs) in external
knowledge, yet the design space of retrieval pipelines is large and the
trade-offs between variants are not well understood, especially on
domain-specific corpora at realistic scale. In this work, we present a
controlled comparison of six retrieval strategies for scientific question
answering: (i) classic top-k dense retrieval, (ii) LLM-based query rephrasing,
(iii) query rephrasing followed by LLM-based reranking, (iv) multi-query fusion via Reciprocal Rank Fusion (RRF), (v) an agentic tool-call pipeline in which the generator decides for itself whether to retrieve, and (vi) late-interaction retrieval with ColBERTv2. All six pipelines share the same generator (Meta-Llama/Llama-3.1-8B-Instruct), prompt, and evaluation protocol; the five single-vector pipelines additionally share SPECTER2 embeddings and a Chroma vector store; and all six retrieve from the full corpus of 463,971 arXiv papers dated 2024--2025. To support
reproducible, large-scale evaluation, we also release a synthetic question
dataset of 19,484 problem-statement and methodology questions generated by
Llama-3.1-8B-Instruct from a random sample of 10,000 papers across academic
domains (query generation succeeded for 9,742 of them), and every strategy is
evaluated on this same query set. We describe the architecture and
implementation of each pipeline, release the code and the synthetic question dataset, and evaluate each strategy with an LLM-as-a-judge protocol
along multiple quality dimensions, together with direct gold-paper retrieval
metrics. The result is an open testbed for studying the cost and quality
trade-offs of RAG design choices on a research-literature corpus, and a basis
for future work on faithfulness, retrieval robustness, and agentic
retrieval.}


\section{Introduction and Motivation}

Large Language Models (LLMs) have become highly effective for natural language
understanding and generation; however, their reliance on static parametric
memory makes them prone to hallucinations, outdated information, and
unsupported factual claims, particularly in scientific and technical domains
[9], [10]. Retrieval-Augmented Generation (RAG), introduced by Lewis et
al.~[12], addresses this limitation by coupling language models with external
retrieval systems that provide dynamically retrieved evidence during
generation.

In a standard RAG pipeline, user queries are embedded into vector
representations, relevant documents are retrieved from a vector store, and the
retrieved context is incorporated into the model prompt for response
generation. Although conceptually simple, the effectiveness of RAG systems
heavily depends on retrieval quality, ranking strategies, query formulation,
and context selection mechanisms [5, 7].

Recent work has proposed multiple enhancements to traditional retrieval
pipelines, including query rewriting [14], hypothetical document generation
[6], reranking approaches [16], reciprocal rank fusion methods [3],
self-reflective retrieval systems such as Self-RAG [1], and agentic retrieval
frameworks [15], [18], [19]. While these approaches improve retrieval quality
and grounding, they also introduce additional computational cost and
architectural complexity.

Despite rapid progress in Retrieval-Augmented Generation research,
practitioners still lack clear guidance regarding which retrieval strategies
perform best under realistic deployment settings and large-scale scientific
corpora. Existing studies often evaluate isolated retrieval components or rely
on small benchmark datasets that may not reflect real-world scientific search
environments.

This paper addresses these limitations by constructing a unified evaluation
framework for comparing retrieval pipelines in scientific question answering
systems. We evaluate six retrieval strategies: classic dense retrieval,
LLM-based query rephrasing, query rephrasing with reranking, multi-query fusion using Reciprocal Rank Fusion (RRF), agentic retrieval, and late-interaction retrieval with ColBERTv2. All pipelines share a common corpus, generator model, and evaluation setup, and the five single-vector pipelines share a common embedding model and vector database, to ensure fair comparison.

The contributions of this work are summarized as follows:
\begin{enumerate}
  \item We construct a unified RAG evaluation framework over a large-scale
        corpus of scientific literature.
  \item We compare six retrieval pipeline strategies under controlled
        experimental settings.
  \item We release a synthetic scientific question-answering dataset generated
        from arXiv papers across multiple domains.
  \item We analyze retrieval quality and response generation using
        LLM-as-a-judge evaluation methodologies and retrieval-oriented
        metrics.
\end{enumerate}

The remainder of this paper is organized as follows. Section~2 reviews related
work on RAG systems and retrieval optimization
techniques. Section~3 presents the methodology and retrieval pipelines,
together with the datasets and evaluation setup. Section~4 reports
experimental results and analysis. Section~5 discusses the findings,
Section~6 outlines limitations, followed by conclusions in Section~7.

\section{Related Work}

\subsection{Retrieval-Augmented Generation}

Retrieval-Augmented Generation (RAG), introduced by Lewis et al.~[12],
combines parametric language models with external retrieval mechanisms to
improve factual grounding and incorporate dynamically retrieved knowledge
during generation. Their work established the retrieve-then-read framework
that forms the foundation of most modern RAG systems.

Closely related work includes REALM [8], which jointly pre-trains retrieval
and language modeling components, and Dense Passage Retrieval (DPR) [11],
which demonstrated the effectiveness of dense dual-encoder retrieval methods
for open-domain question answering. Subsequent surveys organized RAG systems
into pre-retrieval, retrieval, and post-retrieval stages while highlighting
challenges related to retrieval quality, ranking, and context utilization
[5], [7]. For scientific text specifically, SPECTER [24] and its successor
SPECTER2 [25] learn document-level embeddings from citation signals and
provide task-specific adapters for retrieval.

\subsection{Query Reformulation and Expansion}

A common limitation in retrieval systems is that user queries are often short,
ambiguous, or poorly aligned with the target corpus. Query rewriting methods
aim to improve retrieval effectiveness by reformulating queries before
retrieval.

Ma et al.\ introduced query rewriting approaches for Retrieval-Augmented
Generation systems, demonstrating improvements in downstream retrieval and
generation quality [14]. HyDE proposed generating hypothetical documents to
bridge semantic gaps between user queries and retrieved documents [6].

Recent work extended these ideas toward multi-query and decomposition-based
retrieval methods. RQ-RAG introduced learned query refinement mechanisms for
retrieval optimization [2], while DMQR-RAG proposed diverse multi-query
rewriting strategies to improve retrieval coverage and robustness [13].

\subsection{Reranking Methods}

Initial dense retrieval rankings are often imperfect because embedding
similarity does not always correlate with true document relevance. Reranking
approaches address this limitation by re-evaluating retrieved candidates using
more computationally expensive models.

Passage reranking using BERT-based cross-encoders demonstrated substantial
improvements in top-k retrieval precision [16]. Subsequent work extended
reranking approaches to sequence-to-sequence ranking architectures for
improved contextual ranking performance [17].

Two-stage retrieval pipelines consisting of retrieval followed by reranking
have therefore become common in production RAG systems.

\subsection{Rank Fusion Techniques}

When multiple retrieval strategies or multiple query variants are available,
rank fusion techniques provide mechanisms for combining ranked retrieval
outputs.

Reciprocal Rank Fusion (RRF), introduced by Cormack et al.~[3], aggregates
rankings from multiple retrieval systems using reciprocal ranking scores and
has been widely adopted in hybrid and multi-query retrieval pipelines due to
its simplicity and effectiveness. 
Recent RAG systems frequently employ fusion techniques to improve retrieval
diversity and robustness across heterogeneous retrieval outputs.

\subsection{Agentic and Tool-Augmented Retrieval}

Recent research has explored agentic retrieval systems in which language
models dynamically decide when and how to retrieve information during
generation.

ReAct introduced reasoning-and-acting paradigms that interleave
chain-of-thought reasoning with external tool usage [19]. Toolformer further
demonstrated that language models can learn to invoke external tools
autonomously [18]. Self-RAG extended retrieval systems by jointly learning
retrieval, generation, and critique mechanisms through self-reflection [1].
More recent frameworks, such as MA-RAG, introduced collaborative multi-agent
retrieval for improving reasoning and evidence aggregation [15].

\subsection{Evaluation of RAG Systems}

Evaluating RAG systems remains challenging because traditional lexical overlap
metrics such as BLEU and ROUGE do not adequately capture factual grounding,
faithfulness, or retrieval relevance. Jiang et al.\ found LLMs fail to
recognize the personal preferences of user profiles across time [21]. Recent
work introduced LLM-as-a-judge methodologies capable of approximating human
preference evaluation for open-ended tasks [20]. RAG-specific evaluation
frameworks such as RAGAS further extended evaluation toward faithfulness,
answer relevance, and contextual grounding [4]. These evaluation methods have
become increasingly important for assessing retrieval quality and generation
reliability in large-scale RAG systems.

Despite rapid progress in RAG research, limited
work systematically compares multiple retrieval pipeline strategies under
controlled conditions using large-scale scientific corpora. Existing studies
often evaluate isolated retrieval components or focus on small benchmark
datasets. Our work addresses this gap by benchmarking dense retrieval, query rewriting, reranking, rank fusion, agentic retrieval, and late-interaction retrieval approaches within a unified evaluation framework for scientific
question answering systems.

\section{Methods}

\subsection{Dataset, setup, synthetic query generation}

To evaluate retrieval performance, we constructed a synthetic scientific
question-answering dataset derived from recent arXiv publications. We used the arXiv metadata mirror released on Zenodo (record 15808027, $\sim$4.7~GB, $\sim$1.7M records) and retained the 463,971 records whose metadata date falls in \textit{2024 or 2025}. This date is the record's last-updated date rather than its first submission date, so some retained papers were first submitted earlier (1,928 of the 9,742 gold papers, 19.8\%, carry pre-2024 arXiv identifiers). Our synthetic dataset was generated from a random sample of 10,000 of these papers across
multiple academic domains including computer science, mathematics, physics,
biology, and engineering. Each dataset instance contains the paper title,
abstract, publication date, category labels, and a unique arXiv identifier. We
draw the random sample of 10,000 papers using single-pass reservoir sampling
(seed 42) over the streamed JSON.

The questions dataset was developed synthetically using
Meta-Llama/Llama-3.1-8B-Instruct. For each sampled paper, the generator model
produces two synthetic search queries from the abstract, designed to retrieve
that specific paper. The two retrieval-oriented queries generated per paper focus on different aspects of the research contribution:

\begin{enumerate}
  \item \textbf{Problem Query} focused on the problem / research gap the paper
        addresses. It captures the central challenge, limitation, or research
        problem addressed by the paper.
  \item \textbf{Method Query} on the methodology / technical approach in the
        paper. It captures the methodology, architecture, algorithm, or
        experimental approach.
\end{enumerate}

Queries are constrained to 1--2 lines, written in natural language, and
prohibited from reusing the paper title verbatim. Question generation succeeded for 9,742 of the 10,000 papers (258 were dropped because the model output could not be parsed as JSON, the request failed, or the record had no abstract), yielding \textit{19,484 query--gold-paper pairs} (two queries per paper: a problem and a method query) used to evaluate every retrieval strategy. The paper that a
query was generated from serves as its gold target. Some examples of generated
queries include:

\medskip
\noindent\textbf{Machine Learning}
\begin{itemize}[label={-}]
  \item Problem Query: \textit{How do existing click-through rate prediction
        models fail to capture complex feature relationships in large-scale
        industrial data?}
  \item Method Query: \textit{What is a novel approach to combining multiple
        neural network branches for improved feature interaction modeling and
        generalization?}
\end{itemize}

\noindent\textbf{Natural Language Processing}
\begin{itemize}[label={-}]
  \item Problem Query: \textit{How do English-centric training corpora impact
        multilingual capabilities in large language models?}
  \item Method Query: \textit{What novel fine-tuning paradigms establish
        cross-lingual connections at the latent level for multilingual
        language models?}
\end{itemize}

\noindent\textbf{Knowledge Graphs}
\begin{itemize}[label={-}]
  \item Problem Query: \textit{How do path-based models address the sparseness
        problem in knowledge graphs?}
  \item Method Query: \textit{What is an alternative approach to using
        external models for path reasoning in sparse knowledge graphs?}
\end{itemize}

Given a generated query, the retrieval objective is to identify the
corresponding source paper together with semantically related documents from
the corpus. This design allows evaluation of lexical and semantic retrieval
performance while providing a realistic benchmark for retrieval pipelines.

\subsubsection*{Embeddings and vector store}

Papers are converted to embeddings and loaded into a persistent \textit{Chroma}
vector collection (\texttt{arxiv\_papers}). Each document is the concatenation
of the paper title, abstract, and category labels, formatted as:

\begin{quote}
\small\ttfamily
Title: \{title\}\\
Abstract: \{abstract\}\\
Categories: \{categories\}
\end{quote}

\noindent with submitter, categories, and date retained as metadata.
Embeddings are produced with the \textbf{SPECTER2} scientific-document model
(\texttt{allenai/specter2\_base}): documents are encoded with the proximity
adapter and queries with the ad-hoc query adapter, with a maximum sequence
length of 512 tokens, running on GPU. The vector store is queried with cosine
distance, retrieving the top $k=3$ documents per query unless stated otherwise.

\subsection{Models and serving}

These are the models and explanations on their utility:

\begin{itemize}[label={-}]
  \item \textit{Generator:} Llama-3.1-8B-Instruct (4-bit Q4\_K\_M
        quantization), served locally with Ollama (OpenAI-compatible endpoint,
        \texttt{localhost:11434}). Generation uses temperature 0; for prompt-completion calls an output-token budget is computed per prompt from an assumed 8192-token context window ($\sim$4 characters per token, 300-token safety buffer) to avoid context overflow.
  \item \textit{Judge:} Qwen2.5-32B-Instruct (Ollama tag \texttt{qwen2.5:32b}, Q4\_K\_M quantization, served on the same Ollama endpoint) used for the LLM-as-a-judge scoring
        described in \S3.4. Judging is performed in parallel across worker
        threads.
  \item \textit{ColBERT retriever:} ColBERTv2 with the PLAID index [22, 23],
        indexed over the full corpus with a document length limit of 512 tokens, a query length limit of 64 tokens, and 2-bit residual compression.
\end{itemize}

The judge does not share a model family with the generator,
reducing the risk of self-preference bias. All experiments were
run on a single NVIDIA DGX Spark (GB10 Grace Blackwell
superchip, 128 GB unified memory).

\subsection{Retrieval strategies}

All six strategies share the same generator and the same top-$k=3$ documents passed to the generator, and five of the six share the same answer-grounding prompt template (answer only from retrieved papers; do not use outside knowledge; state when evidence is insufficient); Tool Call RAG uses a similar system-prompt formulation (answer only from retrieved papers when retrieving; say the information is insufficient otherwise) because its answer is produced inside a tool-calling chat turn. They differ only in \textit{how} those documents are retrieved. The first five strategies (Classic, Query Rephrased, Rephrased \& Reranked, Fusion, and Tool Call) retrieve from the Chroma vector store described in \S3.1 and are referred to below as the Chroma-based strategies; ColBERT RAG retrieves from its own PLAID index.

\begin{enumerate}
  \item \textit{Classic RAG (baseline).} Embed the user query, retrieve the
        top-3 papers from Chroma, answer directly from them.
  \item \textit{Query Rephrased RAG.} The generator first rewrites the
        question into a clean academic semantic-search query (preserving the original technical terms and adding two to five related academic terms, without Boolean or search-engine syntax); the rewritten
        query drives the vector search.
  \item \textit{Rephrased \& Reranked RAG.} Query rewriting as above, followed by an LLM re-ranking stage: a larger candidate set (10 documents) is retrieved and re-scored listwise by the generator itself, and only the best three documents after re-ranking are passed to the generator.
  \item \textit{Fusion RAG (RAG-Fusion / RRF).} The generator produces three diverse sub-queries for the same question; each is searched independently (five hits each) and the ranked lists are merged with \textit{Reciprocal Rank Fusion} ($k=60$) into a single fused ranking; the original query is searched only as a fallback when the model returns fewer than three usable sub-queries (46 of 19,484 queries).
  \item \textit{Tool Call RAG (agentic).} Retrieval is exposed to the model as a callable tool, so the model decides whether to retrieve rather than always retrieving up front; at most one search is executed (single-step, non-iterative), after which the model answers from its result.
  \item \textit{ColBERT RAG (late-interaction dense retrieval).} In late-interaction retrieval, each query and each document is represented by one embedding per token rather than a single pooled vector, and relevance is scored by matching every query token against its most similar document token and summing the results (the \textit{MaxSim} operator) [22, 23]. Because the document-side embeddings are precomputed, this fine-grained matching is deferred until query time, hence \textit{late} interaction. We index the full 463,971-paper corpus (the same document pool available to the Chroma-based strategies) with \textit{PLAID}, using a document length limit of 512 tokens, and retrieve with MaxSim over these token-level embeddings. The answer-generation
        loop is otherwise identical to the other strategies.
\end{enumerate}

Each of the above strategies writes its generated answers to
\texttt{arxiv\_2025\_llama\_8b\_I\_<strategy>\_rag.jsonl}.

\subsection{LLM-as-a-judge evaluation}

Every generated answer is scored by the judge model against its originating
question. The judge sees only the question and the answer, not the retrieved
papers, so every rating is made from the answer text alone. For each (\texttt{question, answer}) pair the judge first decides
\texttt{is\_answer} --- whether the answer makes a genuine attempt to answer
the question (as opposed to refusing, claiming insufficient context, or
returning something empty/irrelevant). If \texttt{is\_answer} is false, all
metric scores are set to 0; otherwise, the judge assigns an integer
\textit{1--5} rating on each of:

\begin{itemize}[label={-}]
  \item \textit{Accuracy}: apparent factual correctness of the answer,
  \item \textit{Completeness}: coverage of the question,
  \item \textit{Faithfulness}: absence of fabricated or unsupported claims,
        as far as this can be judged from the answer text,
  \item \textit{Relevance}: on-topic-ness of the answer,
  \item \textit{Clarity}: readability and coherence,
\end{itemize}

plus an \textit{Overall} score. The prompt names these six metrics but does not define them, so the judge scores from the metric names alone; the descriptions above are our intended reading of each score. Scores per strategy are written to a file
\texttt{*\_rag\_eval.jsonl} in parallel. For example, the judge decided
\texttt{is\_answer==false} and set all scores to 0 for pair (``What
conditions on subgroups lead to exotic C$^*$-completions of double
groupoids?'', ``The available information is insufficient to determine the
conditions on subgroups that lead to exotic C$^*$-completions of double
groupoids.''); and \texttt{is\_answer==true} for (``How do stellar magnetic
fields in partly and fully convective stars exhibit dichotomous behavior?'',
``Stellar magnetic fields in partly and fully convective stars exhibit
dichotomous behavior, with some stars showing strong, mostly axisymmetric,
and dipole-dominated magnetic fields, while others display weak,
non-axisymmetric, and multipole-dominated fields\ldots'').

\subsection{Aggregation}

For each strategy we report results separately for the \textit{problem} and
\textit{method} query subsets. We report (i) the \textit{answer rate} --- the
percentage of queries with \texttt{is\_answer = true} --- and (ii) the mean of
each 1--5 metric. By convention, \textit{zero-valued scores are excluded} from the metric means, so per-metric averages reflect
quality \textit{conditional on the model attempting an answer}, while the
answer rate captures how often it attempts one. We additionally report
\textit{unconditional} means, in which refusals remain as zeros, providing a
single end-to-end score over all queries. All six strategies were run on the same 9,742 papers for which query generation succeeded, so each strategy has 19,484 question instances and every cross-strategy comparison is paired at the query level. Because every synthetic query has a known gold source paper, we
also compute retrieval metrics directly from the run files: Hit@1 and Hit@3
measure whether the gold paper appears at those cutoffs, and MRR@3 rewards
higher gold-paper rank.

\section{Results}

\subsection{Overall comparison}

Judge scores span 3.47--3.96 across strategies and query types, with
completeness the lowest-scoring dimension and relevance the most
discriminative across strategies. Table~\ref{tab:overall} summarizes overall answer quality across the six
strategies. \textit{ColBERT RAG achieves the highest overall judge score on
both query types} (mean overall 3.94/5), ahead of \textit{Rephrased \&
Reranked RAG} (3.77) and the \textit{Classic RAG} baseline (3.60). Among the
Chroma-based strategies, Rephrased \& Reranked is the strongest, while
\textit{Fusion} (3.55) and \textit{Tool Call} (3.54) fall below the Classic
baseline. The ranking is unchanged under the unconditional measure (refusals
kept as zeros; last column of Table~\ref{tab:overall}: ColBERT 3.94, Rephrased \& Reranked 3.75), so it is not an
artifact of excluding refusals.

\begin{table}[t]
\centering
\caption{Overall judge score (1--5) by strategy: conditional means (zeros excluded) and, in the last column, the unconditional mean with refusals kept as zeros.}
\label{tab:overall}
\footnotesize
\setlength{\tabcolsep}{4pt}
\begin{tabular}{lcccc}
\toprule
RAG Strategy & \makecell[c]{Problem\\overall} & \makecell[c]{Method\\overall} & \makecell[c]{Mean\\overall} & \makecell[c]{Mean overall\\(uncond.)} \\
\midrule
Classic              & 3.71 & 3.50 & 3.60 & 3.57 \\
Query Rephrased      & 3.69 & 3.55 & 3.62 & 3.59 \\
Rephrased\&Reranked  & 3.84 & 3.71 & 3.77 & 3.75 \\
Fusion               & 3.64 & 3.47 & 3.55 & 3.49 \\
Tool Call            & 3.57 & 3.51 & 3.54 & 3.46 \\
\textbf{ColBERT}     & \textbf{3.96} & \textbf{3.92} & \textbf{3.94} & \textbf{3.94} \\
\bottomrule
\end{tabular}
\end{table}

\subsection{Answer rate}

All six strategies answer at least 96.9\% of queries. \textit{ColBERT attains
the highest answer rates} (99.9\% on both query types), while \textit{Tool Call}
is lowest (98.3\% problem, 96.9\% method). 
The Tool Call agentic model retrieves in all but 57 of its 19,484 queries. Of the 57 queries where the agent skipped retrieval, in 30 the agent answered from parametric knowledge, with no retrieved evidence, and the judge (which never sees evidence) accepted them as genuine answers; 
there were 27 other cases where the agent skipped retrieval and the judge scored the outcome as a non-answer. 
Out of 467 refusals to answer in total, 440 followed a retrieval but the model declared the retrieved evidence insufficient.

\begin{table}[t]
\centering
\caption{Answer rate (\texttt{is\_answer = true}) by strategy and query type
(problem vs.\ method).}
\label{tab:answerrate}
\small
\begin{tabular}{lcc}
\toprule
RAG Strategy & \makecell[c]{Problem-query\\answer rate} & \makecell[c]{Method-query\\answer rate} \\
\midrule
Classic              & 99.2\% & 98.8\% \\
Query Rephrased      & 99.1\% & 99.0\% \\
Rephrased\&Reranked  & 99.3\% & 99.3\% \\
Fusion               & 98.7\% & 97.9\% \\
Tool Call            & 98.3\% & 96.9\% \\
\textbf{ColBERT}     & \textbf{99.9\%} & \textbf{99.9\%} \\
\bottomrule
\end{tabular}
\end{table}

\subsection{Per-metric scores}

Tables~\ref{tab:problem} and~\ref{tab:method} give the full per-metric
breakdown for the problem and method query subsets. Three consistent patterns
emerge:

\begin{itemize}[label={-}]
  \item \textit{ColBERT leads on essentially every metric.} It is the top
        scorer on accuracy, completeness, faithfulness, and relevance for both
        query types, with the largest margins on relevance (4.86 problem / 4.81 method) and accuracy (3.93 / 3.90) --- the two metrics most directly tied to retrieving the right evidence.
  \item \textit{LLM re-ranking remains the best single-vector pipeline.}
        Rephrased \& Reranked is the top Chroma-based strategy on accuracy, faithfulness, relevance, and overall for both query types (Tool Call edges it on completeness and clarity), confirming that adding a model-based re-ranking
        pass on top of query rewriting improves the final evidence set. Query
        rewriting alone is still not enough: Query Rephrased trails Classic on
        problem queries.
  \item \textit{Completeness is the binding constraint.} Completeness is the lowest-scoring metric for every strategy (2.71--3.17), while relevance and faithfulness are the highest.
\end{itemize}

\begin{table}[t]
\centering
\caption{Mean per-metric judge scores (1--5, zeros excluded) --- PROBLEM
queries.}
\label{tab:problem}
\footnotesize
\setlength{\tabcolsep}{3.2pt}
\begin{tabular}{lcccccc}
\toprule
RAG & \makecell[c]{Accu\\racy} & \makecell[c]{Compl\\eteness} & \makecell[c]{Faithf\\ulness} & \makecell[c]{Relev\\ance} & \makecell[c]{Clar\\ity} & \makecell[c]{Over\\all} \\
\midrule
Class.        & 3.64 & 2.95 & 4.41 & 4.53 & 3.91 & 3.71 \\
Rephr.        & 3.61 & 2.94 & 4.39 & 4.49 & 3.91 & 3.69 \\
R \& R        & 3.77 & 3.10 & 4.56 & 4.66 & 3.98 & 3.84 \\
Fusion        & 3.55 & 2.85 & 4.32 & 4.45 & 3.88 & 3.64 \\
ToolC         & 3.49 & 3.14 & 4.28 & 4.29 & \textbf{4.05} & 3.57 \\
\textbf{ColB} & \textbf{3.93} & \textbf{3.17} & \textbf{4.71} & \textbf{4.86} & 4.05 & \textbf{3.96} \\
\bottomrule
\end{tabular}
\end{table}

\begin{table}[t]
\centering
\caption{Mean per-metric judge scores (1--5, zeros excluded) --- METHOD
queries.}
\label{tab:method}
\footnotesize
\setlength{\tabcolsep}{3.2pt}
\begin{tabular}{lcccccc}
\toprule
RAG & \makecell[c]{Accu\\racy} & \makecell[c]{Compl\\eteness} & \makecell[c]{Faithf\\ulness} & \makecell[c]{Relev\\ance} & \makecell[c]{Clar\\ity} & \makecell[c]{Over\\all} \\
\midrule
Class.        & 3.39 & 2.75 & 4.03 & 4.21 & 3.79 & 3.50 \\
Rephr.        & 3.44 & 2.80 & 4.09 & 4.26 & 3.84 & 3.55 \\
R \& R        & 3.64 & 2.96 & 4.30 & 4.48 & 3.91 & 3.71 \\
Fusion        & 3.36 & 2.71 & 3.99 & 4.19 & 3.76 & 3.47 \\
ToolC         & 3.39 & 3.04 & 4.05 & 4.06 & 3.99 & 3.51 \\
\textbf{ColB} & \textbf{3.90} & \textbf{3.08} & \textbf{4.54} & \textbf{4.81} & \textbf{4.00} & \textbf{3.92} \\
\bottomrule
\end{tabular}
\end{table}

\subsection{Retrieval effectiveness}

Because every synthetic query has a known gold source paper, we can measure
retrieval quality directly, independent of the judge.
Table~\ref{tab:retrieval} reports gold-paper Hit@1, Hit@3, and MRR@3 per
strategy and query type. \textit{ColBERT retrieves the gold paper within the
top 3 for 92.8\% of problem queries and 94.8\% of method queries} --- roughly
double the best single-vector pipeline (Rephrased \& Reranked: 47.5\% and
52.8\%). Among the Chroma-based strategies the ordering broadly mirrors the judge scores: re-ranking improves over Classic, rewriting alone \textit{hurts}
retrieval on problem queries (Hit@3 39.7\% vs.\ 46.3\% for Classic), and
Fusion is weakest. Retrieval quality translates directly into answer quality:
conditioning on whether the gold paper was retrieved, the mean unconditional overall score is 0.36--1.18 points higher on gold-hit queries than on gold-miss queries for
the Chroma-based strategies.

\begin{table}[t]
\centering
\caption{Gold-paper retrieval metrics by strategy and query type.}
\label{tab:retrieval}
\footnotesize
\setlength{\tabcolsep}{4pt}
\begin{tabular}{llccc}
\toprule
RAG Strategy & Query type & \makecell[c]{Hit@1\\(\%)} & \makecell[c]{Hit@3\\(\%)} & \makecell[c]{MRR@3\\(\%)} \\
\midrule
Classic              & problem & 33.7 & 46.3 & 39.3 \\
Classic              & method  & 35.7 & 46.8 & 40.6 \\
Query Rephrased      & problem & 28.3 & 39.7 & 33.3 \\
Query Rephrased      & method  & 34.8 & 45.9 & 39.7 \\
Rephrased\&Reranked  & problem & 36.7 & 47.5 & 41.6 \\
Rephrased\&Reranked  & method  & 44.3 & 52.8 & 48.2 \\
Fusion               & problem & 24.3 & 38.6 & 30.6 \\
Fusion               & method  & 30.2 & 44.2 & 36.4 \\
Tool Call            & problem & 33.6 & 45.7 & 39.0 \\
Tool Call            & method  & 37.5 & 48.5 & 42.4 \\
\textbf{ColBERT}     & problem & \textbf{86.5} & \textbf{92.8} & \textbf{89.4} \\
\textbf{ColBERT}     & method  & \textbf{90.0} & \textbf{94.8} & \textbf{92.2} \\
\bottomrule
\end{tabular}
\end{table}

\subsection{Paired comparison against the Classic baseline}

All strategies answer the same 19,484 queries, so differences can be tested
with paired per-query statistics on the unconditional overall score.
Table~\ref{tab:paired} reports the mean paired delta versus Classic with 95\%
confidence intervals. Only two strategies significantly outperform the
baseline on both query types: \textit{ColBERT} ($+0.27$ problem, $+0.47$
method; pooled $+0.37$, 95\% CI $[+0.36, +0.38]$) and \textit{Rephrased \&
Reranked} ($+0.13$ problem, $+0.23$ method; pooled $+0.18$, 95\% CI
$[+0.17, +0.19]$). \textit{Fusion and Tool Call are significantly worse than
Classic on both query types}, and Query Rephrased is worse on problem queries
but slightly better on method queries --- the added pipeline complexity still needs a re-ranking stage to pay off.

\begin{table*}[t]
\centering
\caption{Paired per-query difference in unconditional overall score vs.\
Classic (95\% CI), $n=9{,}742$ per cell.}
\label{tab:paired}
\small
\begin{tabular}{llccc}
\toprule
Strategy & \makecell[l]{Query\\type} & \makecell[c]{Mean $\Delta$\\vs Classic} & 95\% CI & \makecell[c]{Win / Tie /\\Loss (\%)} \\
\midrule
Query Rephrased     & problem & $-0.023$ & $[-0.040, -0.006]$ & 17.8 / 63.3 / 18.9 \\
Query Rephrased     & method  & $+0.059$ & $[+0.039, +0.079]$ & 22.9 / 58.2 / 18.9 \\
Rephrased\&Reranked & problem & $+0.126$ & $[+0.110, +0.142]$ & 22.5 / 64.1 / 13.4 \\
Rephrased\&Reranked & method  & $+0.231$ & $[+0.212, +0.251]$ & 29.3 / 55.9 / 14.8 \\
Fusion              & problem & $-0.093$ & $[-0.111, -0.076]$ & 15.7 / 61.8 / 22.5 \\
Fusion              & method  & $-0.057$ & $[-0.078, -0.036]$ & 20.5 / 55.2 / 24.3 \\
Tool Call           & problem & $-0.170$ & $[-0.190, -0.150]$ & 21.6 / 44.7 / 33.7 \\
Tool Call           & method  & $-0.050$ & $[-0.073, -0.027]$ & 29.2 / 39.6 / 31.2 \\
\textbf{ColBERT}    & problem & $\mathbf{+0.272}$ & $[+0.255, +0.290]$ & 27.6 / 61.4 / 11.0 \\
\textbf{ColBERT}    & method  & $\mathbf{+0.467}$ & $[+0.446, +0.487]$ & 37.2 / 52.1 / 10.7 \\
\bottomrule
\end{tabular}
\end{table*}

\subsection{Per-strategy figures}

\begin{figure}[t]
\centering
\includegraphics[width=\columnwidth]{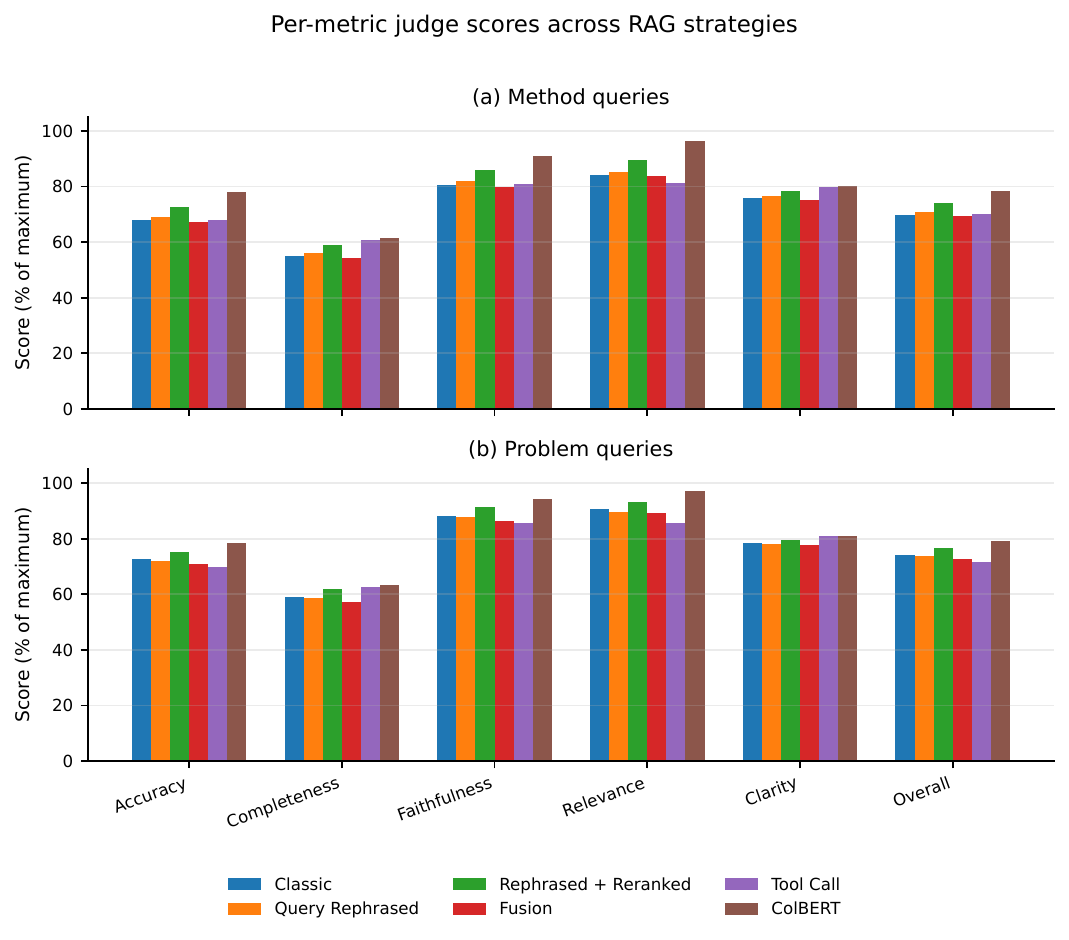}
\caption{The per-metric average score charts for each strategy: Classic RAG
(baseline), Query Rephrased RAG, Rephrased \& Reranked RAG, Fusion RAG (RRF),
Tool Call RAG (agentic), ColBERT RAG (late-interaction).}
\label{fig:permetric}
\end{figure}

\begin{figure}[t]
\centering
\includegraphics[width=\columnwidth]{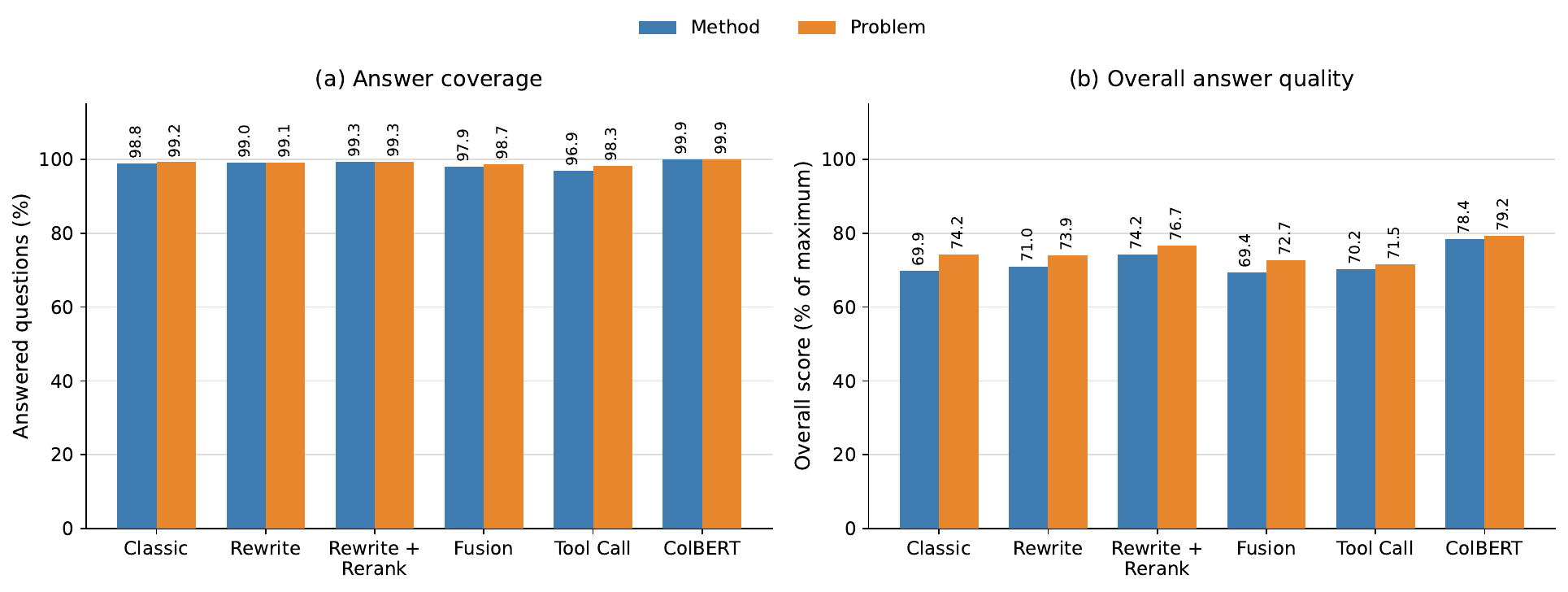}
\caption{The Answer rate (\texttt{is\_answer = true}) and Overall answer
quality by query type (problem vs.\ method) and for each strategy: Classic RAG
(baseline), Query Rephrased RAG, Rephrased \& Reranked RAG, Fusion RAG (RRF),
Tool Call RAG (agentic), ColBERT RAG (late-interaction).}
\label{fig:coverage}
\end{figure}

\begin{figure}[t]
\centering
\includegraphics[width=\columnwidth]{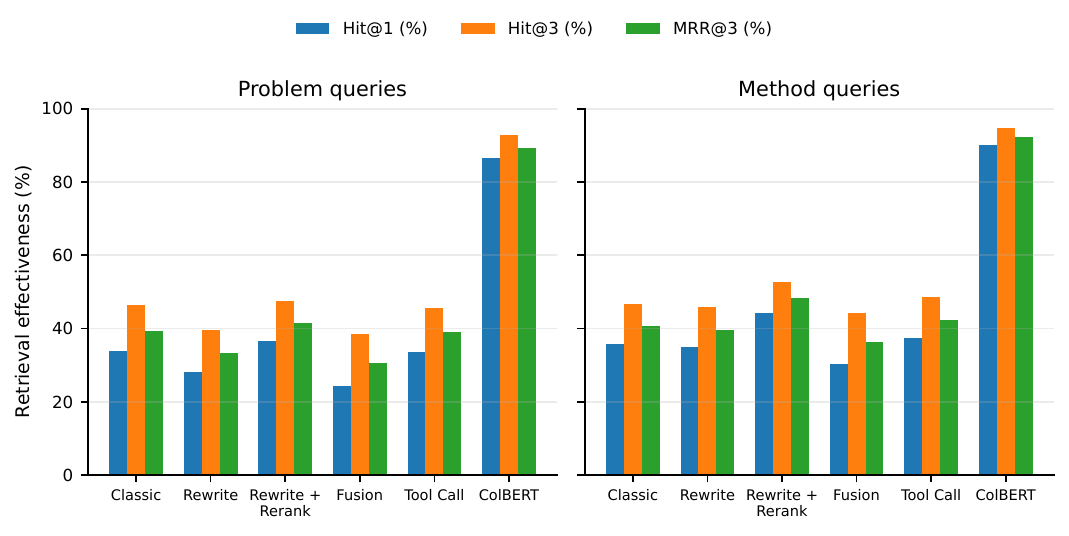}
\caption{Gold-paper retrieval effectiveness (Hit@1, Hit@3, MRR@3) by strategy
and query type.}
\label{fig:retrieval}
\end{figure}

\begin{figure}[t]
\centering
\includegraphics[width=\columnwidth]{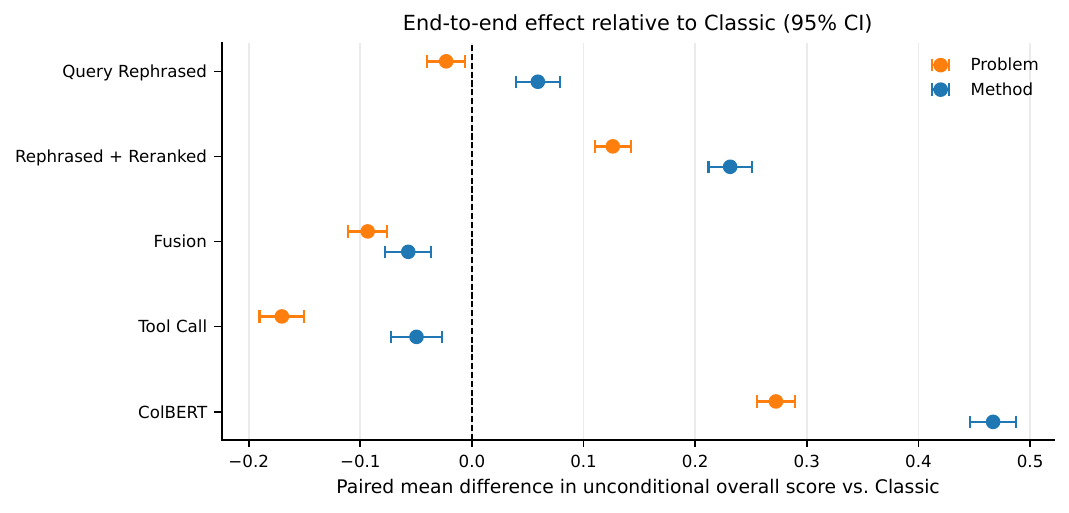}
\caption{Mean paired per-query change in unconditional overall score relative
to the Classic baseline, with 95\% confidence intervals; positive values favor
the named strategy.}
\label{fig:paired}
\end{figure}

Figure~\ref{fig:permetric} shows the per-metric average scores for each RAG
strategy and query type. As shown, ColBERT RAG performed the best across
nearly all metrics, followed by Rephrased \& Reranked RAG; Fusion and Tool
Call trail the Classic baseline. The relative performance stays consistent
across the different metrics. Figure~\ref{fig:coverage} shows the answer rate
and overall answer quality for each RAG strategy and query type, confirming
the conclusions above. Figure~\ref{fig:retrieval} shows gold-paper retrieval
effectiveness per strategy, and Figure~\ref{fig:paired} shows the paired
per-query effect of each strategy relative to Classic with 95\% confidence
intervals.

\subsection{Summary of findings}

\begin{enumerate}
  \item \textit{Late-interaction retrieval is the best pipeline in these
        experiments} (ColBERT, overall 3.94/5), leading on both answer quality
        and gold-paper retrieval (Hit@3 $>$ 92\% for both query types) with
        the highest answer rate (99.9\%).
  \item \textit{Query Rewriting + LLM Re-ranking is the best single-vector
        pipeline} (3.77/5), beating the classic baseline on every metric for
        both query types (pooled paired delta $+0.18$ vs.\ Classic).
  \item \textit{The Classic single-shot baseline remains strong} (3.60/5);
        Fusion (3.55) and agentic Tool Call (3.54) add complexity without
        improving on it --- both are significantly \textit{below} Classic in
        the paired comparison.
  \item \textit{Query Rewriting (without re-ranking) exhibits a lower
        performance} on problem queries (Hit@3 39.7\% vs.\ 46.3\% for
        Classic) --- rewriting without re-ranking can drift away from the
        precise target paper. The re-ranking stage is what recovers and
        improves quality.
  \item \textit{Retrieval quality drives answer quality.} Queries where the gold paper was retrieved score 0.36--1.18 points higher (unconditional overall) than gold misses, and the two best retrievers (ColBERT and Rephrased \& Reranked) are also the two best by judge score; Tool Call is the exception, with Classic-level Hit@3 but the lowest judge score.
  \item \textit{Method questions are harder than problem questions} for every
        strategy --- the problem-minus-method gap ranges from 0.04 (ColBERT) to 0.22 points (Classic; from unrounded means).
\end{enumerate}

\section{Discussion}

The experimental results highlight the critical role of retrieval strategy
selection in RAG pipelines. We observed that pipeline complexity does not
linearly scale with retrieval performance. Among the single-vector strategies,
the ``Rephrased \& Reranked RAG'' approach achieved the highest overall scores
(3.77/5), suggesting that model-based re-ranking effectively mitigates
retrieval noise, while multi-query fusion and the agentic tool-call pipeline
fell below the classic single-shot baseline (3.60/5) despite their additional
LLM calls. This indicates that simpler architectures may offer cost-efficiency
without significantly compromising performance.

The clearest result, however, is the effect of the retrieval architecture
itself. ColBERT's late-interaction retrieval, indexed over the same full
463,971-paper corpus as the Chroma-based strategies, retrieved the gold paper
in the top 3 for 92.8--94.8\% of queries --- roughly double the best
single-vector pipeline --- and produced the best answers on nearly every quality dimension (overall 3.94/5, paired delta $+0.37$ vs.\ Classic). Retrieval quality is linked to index composition, and comparisons of retriever architectures are unreliable unless all retrievers operate over the same document pool.

The gold-hit analysis makes the mechanism explicit: conditional on retrieving
the gold paper, all strategies produce answers of similar quality (overall
$\approx$ 3.8--4.0), so nearly all of the between-strategy differences are
attributable to how often the right evidence reaches the generator, not to how
the generator uses it.

\section{Limitations}

This study has several limitations: (i) The evaluation protocol relies on a
single LLM judge that does not see the retrieved papers, so faithfulness is
assessed from the answer text rather than checked against the evidence;
although the judge (Qwen2.5-32B) is from a different model family and larger
than the generator, LLM-as-a-judge scores remain an imperfect proxy for human
judgment. (ii) Gold relevance is determined by
synthetic queries generated from each paper's abstract, which may capture a
simplified view of research relevance. (iii) Our headline metric aggregation
excludes zero-valued scores, necessitating a joint interpretation of answer
rates and conditional quality scores; we mitigate this by additionally
reporting unconditional means and paired per-query effects. (iv) ColBERT
differs from the other strategies in both its encoder and its late-interaction
scoring, so its advantage cannot be attributed to late interaction alone
versus its underlying representations. (v) All pipelines were tested using a
single 4-bit-quantized 8B-parameter generator at temperature 0, which may
limit the generalizability of these findings to larger or more creative
generation settings. (vi) The synthetic queries were produced by the same
model that answers them (Llama-3.1-8B-Instruct), so the query set defining
gold relevance and the system under test share a model, which may bias retrieval difficulty in ways an independently authored query set would not. (vii) The corpus is filtered on each record's metadata date, which is a last-updated date; about a fifth of the gold papers were first submitted before 2024, so the corpus should be read as papers updated in 2024--2025 rather than strictly published then.

\section{Conclusion}

This work presented a systematic, unified evaluation of retrieval strategies
for scientific question answering. By benchmarking six distinct retrieval
pipelines --- ranging from classic dense retrieval to agentic tool-call
architectures and late-interaction retrieval --- over a corpus of 2024--2025
arXiv papers, we demonstrated that retrieval architecture dominates pipeline
complexity: ColBERT's late-interaction retrieval achieves both the highest
gold-paper retrieval rates and the highest answer quality, while among
single-vector pipelines Rephrased \& Reranked RAG yields the highest answer quality and the simpler Classic baseline provides robust, cost-effective performance.

\section*{Acknowledgements}

The authors thank the Department of Computer Science at San
Jose State University for providing the high-performance computing resources that made the results in this paper possible. 
The authors also thank Aleksander Ershov for locating and providing the arXiv corpus used in this work.

\section*{Code and Data Availability}

\noindent\textsc{Code Repository:}\\
\url{https://github.com/bhagyeshrathi07/rag_eval}

\medskip
\noindent\textsc{Dataset --- random 10,000 sample from arXiv corpus:}\\
\url{https://drive.google.com/file/d/1U1Ut6PHwAMzWo1cM6nvVNNRsu_xz4h2c/view}

\medskip
\noindent\textsc{Synthetic problem query and method query dataset:}\\
\url{https://www.kaggle.com/datasets/bhagyeshrathi/synthetic-problem-query-and-method-query-dataset}

\section*{References}

\begin{enumerate}[label={[\arabic*]},leftmargin=2.4em,itemsep=3pt]
  \item Asai, A., Wu, Z., Wang, Y., Sil, A., and Hajishirzi, H. (2024).
        Self-RAG: Learning to Retrieve, Generate, and Critique through
        Self-Reflection. In Proceedings of ICLR.
  \item Chan, C.-M., et al. (2024). RQ-RAG: Learning to Refine Queries for
        Retrieval Augmented Generation. arXiv preprint arXiv:2404.00610.
  \item Cormack, G. V., Clarke, C. L. A., and B\"uttcher, S. (2009).
        Reciprocal Rank Fusion Outperforms Condorcet and Individual Rank
        Learning Methods. In Proceedings of SIGIR, pages 758--759.
  \item Es, S., James, J., Espinosa-Anke, L., and Schockaert, S. (2024).
        RAGAS: Automated Evaluation of Retrieval Augmented Generation. In
        Proceedings of EACL: System Demonstrations.
  \item Fan, W., Ding, Y., Ning, L., Wang, S., Li, H., Yin, D., Chua, T.-S.,
        and Li, Q. (2024). A Survey on RAG Meeting LLMs: Towards
        Retrieval-Augmented Large Language Models. arXiv preprint
        arXiv:2405.06211.
  \item Gao, L., Ma, X., Lin, J., and Callan, J. (2022). Precise Zero-Shot
        Dense Retrieval without Relevance Labels. arXiv preprint
        arXiv:2212.10496.
  \item Gao, Y., Xiong, Y., Gao, X., Jia, K., Pan, J., Bi, Y., Dai, Y., Sun,
        J., Wang, M., and Wang, H. (2023). Retrieval-Augmented Generation for
        Large Language Models: A Survey. arXiv preprint arXiv:2312.10997.
  \item Guu, K., Lee, K., Tung, Z., Pasupat, P., and Chang, M.-W. (2020).
        REALM: Retrieval-Augmented Language Model Pre-training. In
        Proceedings of ICML.
  \item Huang, L., Yu, W., Ma, W., Zhong, W., Feng, Z., Wang, H., Chen, Q.,
        Peng, W., Feng, X., Qin, B., and Liu, T. (2025). A Survey on
        Hallucination in Large Language Models: Principles, Taxonomy,
        Challenges, and Open Questions. ACM Transactions on Information
        Systems.
  \item Ji, Z., Lee, N., Frieske, R., Yu, T., Su, D., Xu, Y., Ishii, E.,
        Bang, Y. J., Madotto, A., and Fung, P. (2023). Survey of
        Hallucination in Natural Language Generation. ACM Computing Surveys,
        55(12), 1--38.
  \item Karpukhin, V., O\u{g}uz, B., Min, S., Lewis, P., Wu, L., Edunov, S.,
        Chen, D., and Yih, W.-t. (2020). Dense Passage Retrieval for
        Open-Domain Question Answering. In Proceedings of EMNLP.
  \item Lewis, P., Perez, E., Piktus, A., Petroni, F., Karpukhin, V., Goyal,
        N., K\"uttler, H., Lewis, M., Yih, W.-t., Rockt\"aschel, T., Riedel,
        S., and Kiela, D. (2020). Retrieval-Augmented Generation for
        Knowledge-Intensive NLP Tasks. In Advances in Neural Information
        Processing Systems (NeurIPS).
  \item Li, Z., et al. (2024). DMQR-RAG: Diverse Multi-Query Rewriting for
        RAG. arXiv preprint arXiv:2411.13154.
  \item Ma, X., Gong, Y., He, P., Zhao, H., and Duan, N. (2023). Query
        Rewriting in Retrieval-Augmented Large Language Models. In
        Proceedings of EMNLP.
  \item Nguyen, T., et al. (2025). MA-RAG: Multi-Agent Retrieval-Augmented
        Generation via Collaborative Chain-of-Thought Reasoning. arXiv
        preprint arXiv:2505.20096.
  \item Nogueira, R., and Cho, K. (2019). Passage Re-ranking with BERT. arXiv
        preprint arXiv:1901.04085.
  \item Nogueira, R., Jiang, Z., Pradeep, R., and Lin, J. (2020). Document
        Ranking with a Pretrained Sequence-to-Sequence Model. In Findings of
        EMNLP.
  \item Schick, T., Dwivedi-Yu, J., Dess\`i, R., Raileanu, R., Lomeli, M.,
        Zettlemoyer, L., Cancedda, N., and Scialom, T. (2023). Toolformer:
        Language Models Can Teach Themselves to Use Tools. In Advances in
        Neural Information Processing Systems (NeurIPS).
  \item Yao, S., Zhao, J., Yu, D., Du, N., Shafran, I., Narasimhan, K., and
        Cao, Y. (2023). ReAct: Synergizing Reasoning and Acting in Language
        Models. In Proceedings of ICLR.
  \item Zheng, L., Chiang, W.-L., Sheng, Y., Zhuang, S., Wu, Z., Zhuang, Y.,
        Lin, Z., Li, Z., Li, D., Xing, E. P., Zhang, H., Gonzalez, J. E., and
        Stoica, I. (2023). Judging LLM-as-a-Judge with MT-Bench and Chatbot
        Arena. In Advances in Neural Information Processing Systems
        (NeurIPS), Datasets and Benchmarks Track.
  \item Jiang, B., Hao, Z., Cho, Y.-M., Li, B., Yuan, Y., Chen, S., Ungar,
        L., Taylor, C., Roth, D. (2025). Know Me, Respond to Me: Benchmarking
        LLMs for Dynamic User Profiling and Personalized Responses at Scale.
        10.48550/arXiv.2504.14225.
  \item Keshav Santhanam, Omar Khattab, Christopher Potts, and Matei Zaharia.
        2022. PLAID: An Efficient Engine for Late Interaction Retrieval. In
        Proceedings of the 31st ACM International Conference on Information
        \& Knowledge Management (CIKM '22). Association for Computing
        Machinery, New York, NY, USA, 1747--1756.
        \url{https://doi.org/10.48550/arXiv.2205.09707}
  \item O. Khattab and M. Zaharia, ``ColBERT: Efficient and effective passage
        search via contextualized late interaction over BERT,'' in Proc. 43rd
        Int. ACM SIGIR Conf. Res. Develop. Inf. Retrieval (SIGIR), Virtual
        Event, China, 2020, pp. 39--48.
  \item Cohan, A., Feldman, S., Beltagy, I., Downey, D., and Weld, D. S.
        (2020). SPECTER: Document-level Representation Learning using
        Citation-informed Transformers. In Proceedings of ACL.
  \item Singh, A., D'Arcy, M., Cohan, A., Downey, D., and Feldman, S. (2023).
        SciRepEval: A Multi-Format Benchmark for Scientific Document
        Representations. In Proceedings of EMNLP.
\end{enumerate}

\end{document}